\documentclass[11pt,a4paper]{article}
\usepackage[latin1]{inputenc}
\usepackage[english]{babel}

\usepackage{amsfonts}
\usepackage{amsmath}
\usepackage{amssymb}
\usepackage{amsthm}
\usepackage{bm}
\usepackage[round]{natbib}
\usepackage{graphicx}
\usepackage{caption}
\usepackage{subcaption}
\usepackage{float}
\usepackage[parfill]{parskip}
\usepackage{epstopdf}
\usepackage{appendix}
\usepackage{booktabs}
\usepackage{verbatim}
\usepackage{url}

\newcommand{\dd}{{\mathrm d}}

\newcommand{\e}{{\rm e}}
\newcommand{\E}{{\mathbb E}}

\newcommand{\Q}{{\mathbb Q}}

\newcommand{\R}{{\mathbb R}}
\newcommand{\N}{{\mathbb N}}

\newcommand{\Fcal}{{\mathcal F}}
\newcommand{\Gcal}{{\mathcal G}}

\newcommand{\Kcal}{{\mathcal K}}

\makeatletter
\def\thm@space@setup{%
  \thm@preskip=\parskip \thm@postskip=0pt
}
\makeatother

\newtheorem{proposition}{Proposition}[section]
\newtheorem{lemma}[proposition]{Lemma}

\newtheorem{corollary}[proposition]{Corollary}
\newtheorem{remark}[proposition]{Remark}

\title{A lognormal type stochastic volatility model with quadratic drift\footnote{We thank Damien Ackerer for helpful comments.}}
\author{Peter Carr\footnote{New York University, Tandon School of Engineering}\quad and\quad Sander Willems\footnote{\'Ecole Polytechnique F\'ed\'erale de Lausanne (EPFL) and Swiss Finance Institute}}
\date{July 17, 2019}

\begin{document}
\maketitle
\begin{abstract}
This paper presents a novel one-factor stochastic volatility model where the instantaneous volatility of the asset log-return is a diffusion with a quadratic drift and a linear dispersion function. The instantaneous volatility mean reverts around a constant level, with a speed of mean reversion that is affine in the instantaneous volatility level. The steady-state distribution of the instantaneous volatility belongs to the class of Generalized Inverse Gaussian distributions. We show that the quadratic term in the drift is crucial to avoid moment explosions and to preserve the martingale property of the stock price process. Using a conveniently chosen change of measure, we relate the model to the class of polynomial diffusions. This remarkable relation allows us to develop a highly accurate option price approximation technique based on orthogonal polynomial expansions. 
\end{abstract}

\section{Introduction}
\begin{figure}[H]
    \centering
    \begin{subfigure}[b]{0.45\textwidth}
        \includegraphics[width=\textwidth]{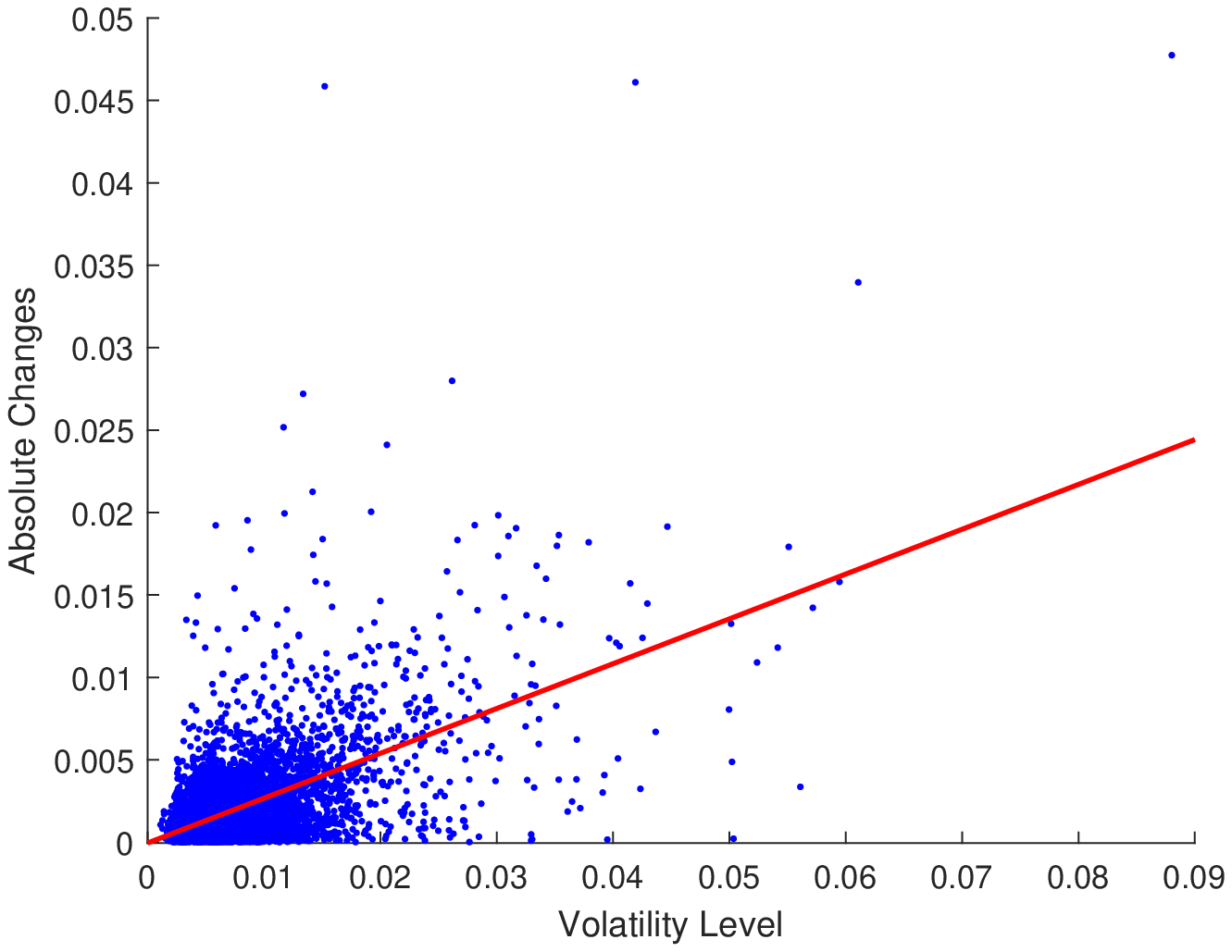}
        \caption{Absolute change in volatility}
        \label{fig:scatter_vol}
    \end{subfigure}
    \begin{subfigure}[b]{0.45\textwidth}
        \includegraphics[width=\textwidth]{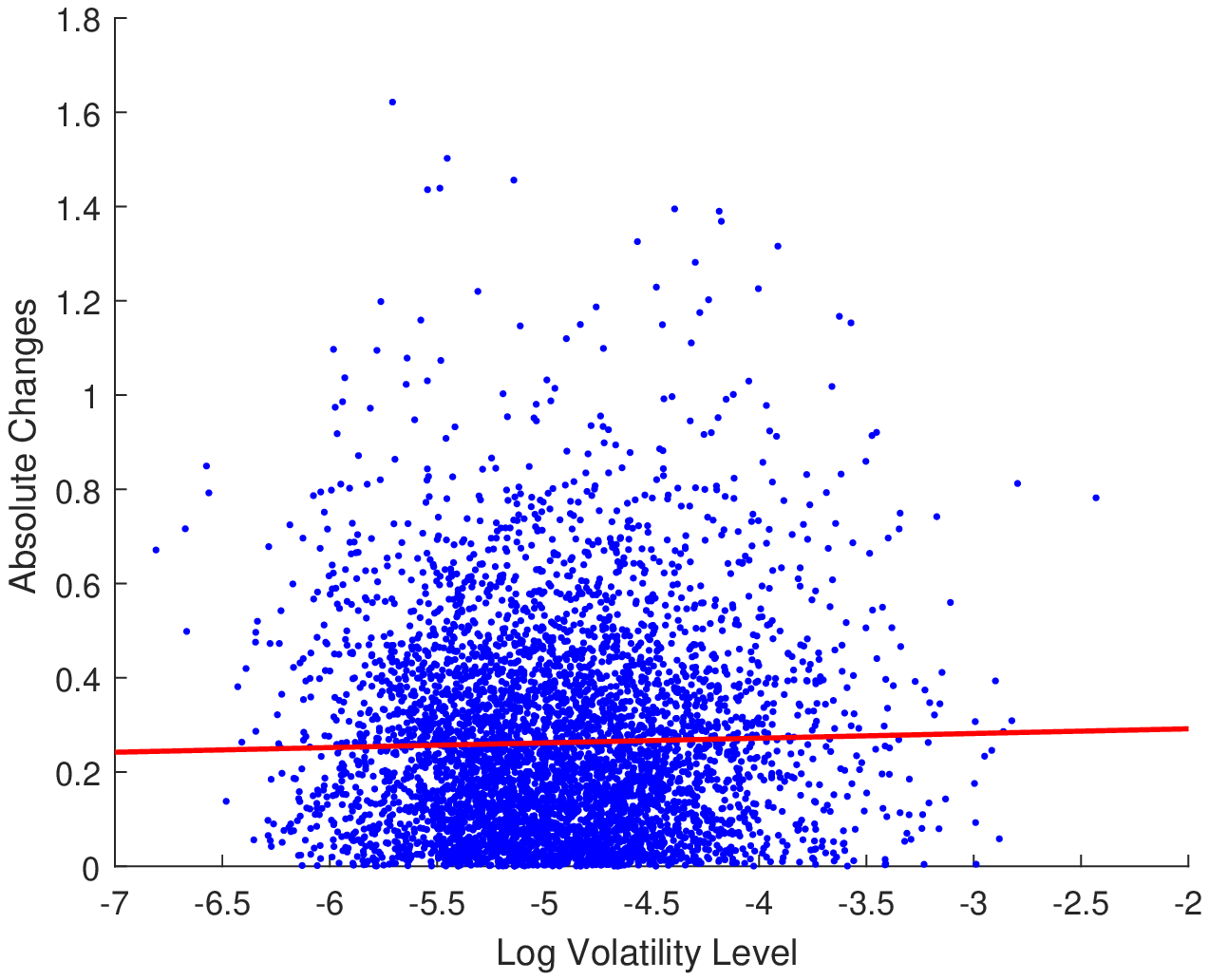}
        \caption{Absolute change in log volatility}
        \label{fig:scatter_logVar}
    \end{subfigure}    
    \begin{subfigure}[b]{0.45\textwidth}
        \includegraphics[width=\textwidth]{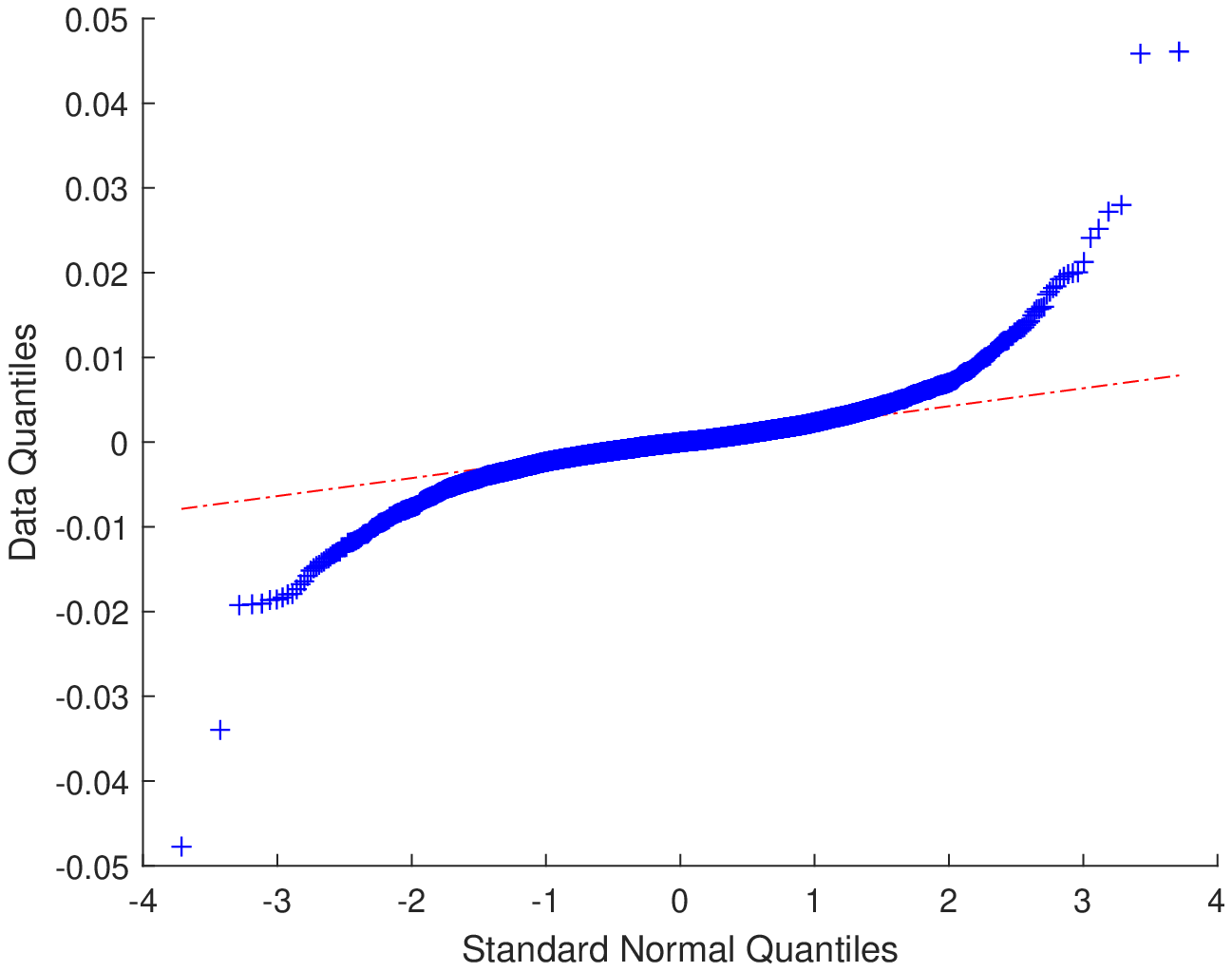}
        \caption{Changes in volatility QQ-plot}
        \label{fig:qqPlot_vol}
    \end{subfigure}
    \begin{subfigure}[b]{0.45\textwidth}
        \includegraphics[width=\textwidth]{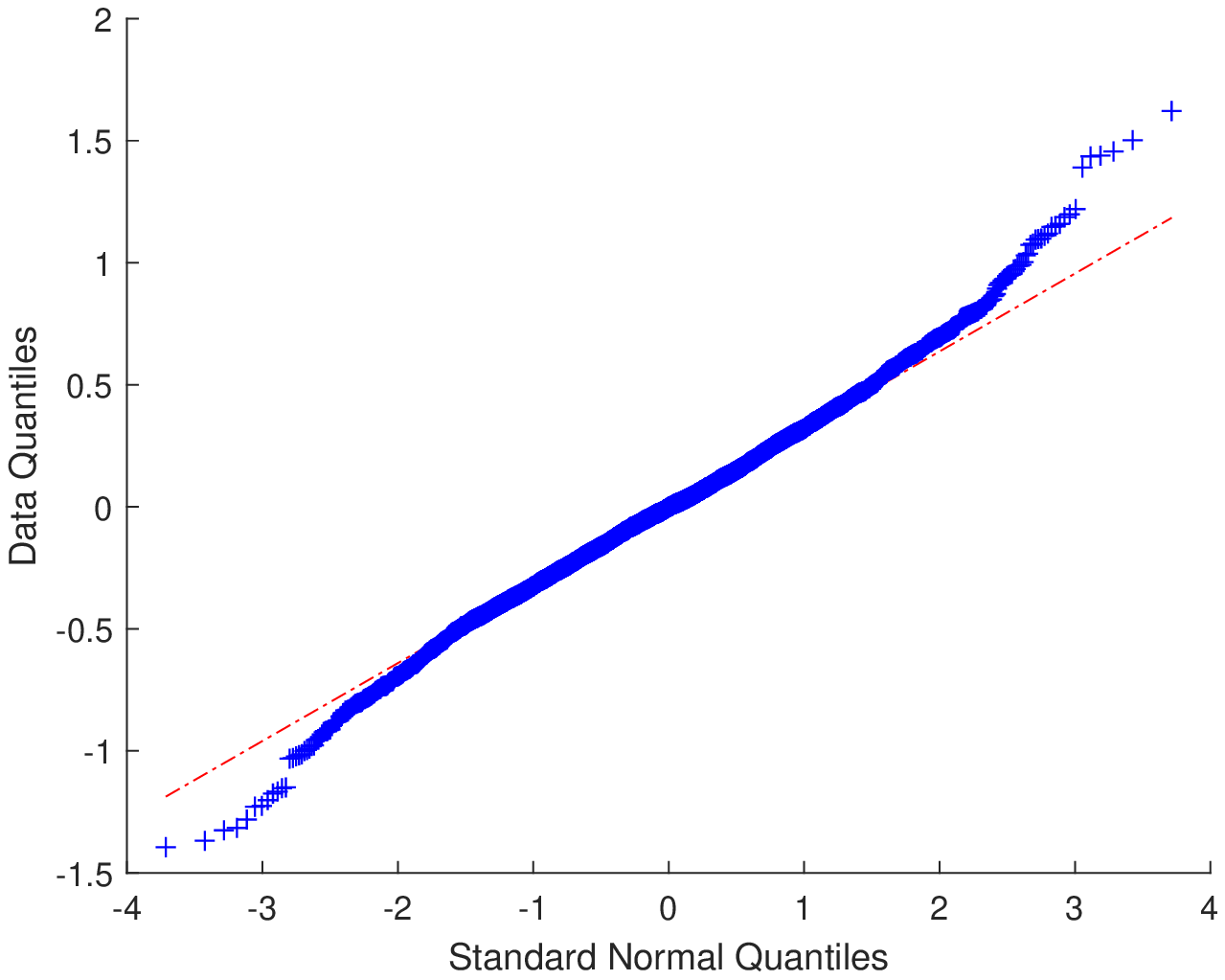}
        \caption{Changes in log volatility QQ-plot}
        \label{fig:qqPlot_logVar}
    \end{subfigure}
\caption{The top left (right) figure shows a scatter plot of the realized (log) volatility level against the absolute change one day ahead, together with a least-squares regression line in red. The bottom left (right) figure shows a quantile-quantile plot of daily changes in realized (log) volatility. Realized volatilities are obtained from Oxford-Man Institute's realized library using 5-minute sub-sampled high-frequency returns on the S\&P500 index from January 2000 until June 2019.}
\label{fig:realized}
\end{figure}

The popularity of the \cite{heston1993closed} model and, more generally, affine models (see e.g., \cite{duffie2003affine}) for modeling stochastic volatility is in large part due to their analytical tractability. However, there is abundant empirical evidence that favours non-affine models, in particular specifications with a lognormal type of diffusion for the (instantaneous) volatility.\footnote{With lognormal type of diffusion we mean a diffusion $\sigma_t$ with $\dd [\sigma,\sigma]_t=\nu^2\sigma_t^2\,\dd t$, for some $\nu>0$.} For instance, \cite{christoffersen2010volatility} show that absolute changes in realized volatility are positively correlated with the volatility level and do not follow a Gaussian distribution. In contrast, the Heston model implies that (instantaneous) changes in volatility should be Gaussian and independent of the volatility level. Changes in the log realized volatility, on the other hand, closely resemble a normal distribution, which motivates the use of a lognormal type of diffusion component in the volatility process.\footnote{See \cite{christoffersen2010volatility} for the S\&P500 index, \cite{andersen2001distributionEQ} for individual stocks in the DJIA index, and \cite{andersen2001distributionFX} for foreign exchange markets.} Figure \ref{fig:realized} reproduces these results using a 5-minute sub-sampled daily realized volatility measure for the S\&P500 index from January 2000 until June 2019 and confirms the findings of \cite{christoffersen2010volatility}.

Lognormal type stochastic volatility models are, however, particularly prone to problems such as moment explosions and loss of the martingale property for the asset price, see e.g.\ \cite{lions2007correlations} and \cite{andersen2007moment}. These problems are caused by the fat right tail of the volatility distribution, which can cause large spikes in the asset price. Having finite higher order moments for the asset price is important, for example, to price derivatives with a super-linear payoff. \cite{andersen2007moment} highlight the importance of these type of derivatives in interest rate markets. Moreover, when pricing derivatives with Monte-Carlo simulations, the payoff needs to have a finite second moment in order to use the central limit theorem to derive confidence intervals on the Monte-Carlo estimators. For example, if the volatility process has an affine drift and a linear dispersion function, then the instantaneous correlation between log-price and volatility has to be smaller than $-87\%$ in order for the asset price to have a finite fourth moment. In this paper, we propose a novel non-affine one-factor stochastic volatility model featuring a diffusion with a quadratic drift function and a linear dispersion function for the volatility process. The quadratic term has a negative coefficient in our model, which allows for a rapid reduction following an upward spike in the volatility.\footnote{\cite{bakshi2006estimation} find empirical evidence for stochastic volatility models with nonlinear drift. In particular, they find a significantly negative coefficient on the quadratic term in the drift.} The linear dispersion function produces lognormal type innovations in the volatility and the quadratic term in the drift controls undesirable side effects such as moment explosions and loss of martingality. Moreover, using the critical moment formula of \cite{lee2004moment}, we show that a nonzero quadratic term in the drift allows to control both the small strike and large strike tail of the Black-Scholes implied volatility skew.  The volatility process in our model is stationary and has a Generalized Inverse Gaussian (GIG) distribution as steady-state distribution. The GIG distribution, which contains the inverse Gaussian, hyperbolic, gamma, and inverse-gamma as special cases, has broad empirical support for modeling stochastic volatility in stock returns, see for example \cite{barndorff1997normal}, \cite{eberlein2001application}, \cite{eberlein2002generalized}, and \cite{gander2007stochastic}.

Since our model is far from affine, tractability is not straightforward. If we set the quadratic term in the drift of the volatility to zero, then our model fits in the class of polynomial diffusions, see e.g.\ \cite{filipovic2016polynomial}. This class of stochastic processes, which contains all affine diffusions as special cases, is characterized by the fact that their infinitesimal generator maps polynomials to polynomials of the same degree or less. As a consequence, all conditional moments of the log-asset price are available in closed form and European style derivatives on the asset price can be priced using moment-based approximation methods, see e.g.\ \cite{ackerer2019option}.\footnote{For applications of polynomial processes in derivative pricing, see for example \cite{filipovic2016quadratic}, \cite{ackerer2016linear}, \cite{filipovic2017term}, \cite{ackerer2018jacobi}, and \cite{willems2019asian}.} The polynomial property is lost, however, as soon as we have a nonzero quadratic term in the drift of the volatility. We circumvent this problem by introducing a change of measure under which the polynomial property is recovered. Under the new measure, derivative prices are given by the expectation of the discounted payoff multiplied by the Radon-Nikodym density of the measure change. We show how to compute all joint conditional moments of the log-asset price and log-Radon-Nikodym density in closed form under the new measure. An orthogonal polynomial expansion technique in the spirit of \cite{ackerer2019option} then allows us to efficiently price European style derivatives on the asset price.


The remaining of this paper is structured as follows. Section \ref{sec:model_spec} describes the model dynamics and Section \ref{sec:steady-state} analyzes the steady-state distribution of the volatility process. In Section \ref{sec:moment_explosions} we study the problem of moment explosions. Section \ref{sec:polynomial_disguise} relates our model to the class of polynomial diffusions, which is used in Section \ref{sec:option_pricing} to develop a derivative pricing approximation method. Section \ref{sec:numerical} contains a numerical study of the model and Section \ref{sec:conclusion} concludes. All proofs and additional technical results are collected in the Appendix.

\section{Model specification} \label{sec:model_spec}
We consider a financial market modeled on a filtered probability space $(\Omega,\Fcal,\Fcal_t,\Q)$, where $\Q$ is a risk-neutral probability measure. Henceforth $\E_t[\cdot]$ denotes the $\Fcal_t$-conditional $\Q$-expectation. Let $S_t$ denote the stock price and assume for simplicity zero interest rates and no dividend payments.\footnote{Alternatively, one can also think of $S_t$ as, for example, an interest rate variable (e.g., forward rate or swap rate) and replace $\Q$ by the appropriate pricing measure.} We specify the following $\Q$-dynamics for the log-price  $x_t=\log(S_t)$ 
\begin{align}
\dd x_t&=-\frac{1}{2}\sigma_t^2\,\dd t+\sigma_t(\rho\,\dd W_t +\sqrt{1-\rho^2}\,\dd B_t), \label{eq:SDE_x}\\
\dd \sigma_t&=(R_0+R_1\sigma_t)(R_2-\sigma_t)\,\dd t + \nu \sigma_t \,\dd W_t,\label{eq:SDE_v}
\end{align}
with $R_0,R_1\ge 0$, $R_2,\nu,\sigma_0>0$, $\rho\in[-1,1]$, $x_0\in\R$, and $W_t,B_t$ independent $\Q$-Brownian motions. The volatility process mean-reverts around a constant level $R_2$ with a stochastic speed of mean-reversion $R_0+R_1\sigma_t$. Our model specification nests many existing models. The lognormal SABR model of \cite{hagan2002managing} arises when  $R_0=R_1=0$, in which case $\sigma_t$ is simply a geometric Brownian motion without drift. In this case, however, the volatility process is not mean-reverting, which is an important empirical feature. If we set $R_0>0$ and $R_1=0$, then $\sigma_t$ becomes a mean-reverting diffusion with affine drift and linear dispersion function, which we refer to as a linear diffusion. This type of model has been studied in \cite{alan2000option}, \cite{sepp2012beta}, \cite{sepp2014empirical,sepp2016log}, \cite{lee2016switching}, and \cite{ackerer2019option}.\footnote{Also related is the GARCH diffusion model of \cite{nelson1990arch} and \cite{barone2005option}, where $\sigma_t^2$ is modeled as a diffusion with affine drift and linear dispersion function. Applying It\^o's lemma shows that the corresponding volatility process also has a linear dispersion function, but it does not have an affine drift. Directly modeling volatility seems more intuitive and provides a more natural interpretation for the model parameters.} 
For $R_0=0$ and $R_1>0$, equation \eqref{eq:SDE_v} is known as the logistic diffusion and originated in the context of modeling constrained population growth in biology, see e.g.\ \cite{tuckwell1987logistic}.\footnote{The deterministic version of this SDE was developed in the early 19th century by the  Belgian mathematician Pierre Fran\c ois Verhulst to model population growth.}  In the context of finance, the logistic diffusion has been used, for example, in a general equilibrium model by \cite{merton1975asymptotic} and in a stochastic volatility model by \cite{hull1987pricing} and \cite{lewis2019exact}.\footnote{\cite{hull1987pricing} specify $\dd \sigma_t^2=a(\sigma^\ast-\sigma_t)\sigma^2_t\,\dd t+\xi\sigma^2_t\,\dd W_t$ for some parameters $a,\sigma^\ast, \xi >0$. Applying It\^o's lemma shows that $\sigma_t$ follows a logistic diffusion: $\dd \sigma_t =\frac{a}{2}\sigma_t(\sigma^\ast-\frac{\xi^2}{4}-\sigma_t)\,\dd t+\frac{\xi}{2}\sigma_t\,\dd W_t$.}

The following proposition shows that the model is well defined and that zero is an unattainable boundary for $\sigma_t$.
\begin{proposition}\label{prop:existence_uniqueness}
There exists a unique strong solution $(x_t,\sigma_t)$ of \eqref{eq:SDE_x}-\eqref{eq:SDE_v} taking values in $\R\times (0,\infty)$.
\end{proposition}
From the proof of Proposition \ref{prop:existence_uniqueness}, it becomes clear that the non-negativity assumption $R_1\ge 0$ is crucial in order for \eqref{eq:SDE_v} to have a global solution, cf.\ Remark \ref{remark:explosion}. Indeed, if $R_1<0$, then $\sigma_t$ blows up in finite time.

\begin{remark}
Although zero is a natural lower bound for the volatility process, we can generalize \eqref{eq:SDE_v} by adding a lower bound $\underline{\sigma}\ge 0$ as follows
\[\dd \sigma_t=(R_0+R_1(\sigma_t-\underline{\sigma}))(R_2-(\sigma_t-\underline{\sigma}))\,\dd t + \nu (\sigma_t-\underline{\sigma}) \,\dd W_t,\]
with $\sigma_0>\underline{\sigma}$ and the same restrictions on the other parameters as before. All the results we derive in our paper are easily adjusted to accommodate this generalization.
\end{remark}

\section{Steady-state distribution of $\sigma_t$} \label{sec:steady-state}
In this following proposition, we explicitly derive the steady-state distribution of $\sigma_t$.
\begin{proposition}\label{prop:steady_state}
If either $R_0>0$, or $R_0=0$ and $2R_1R_2>\nu^2$, then the process $\sigma_t$ has a steady-state distribution with density function 
\[
\pi(x)\propto x^{\xi-1}\exp\left\{-2\frac{R_0R_2}{\nu^2}\frac{1}{x}-2\frac{R_1}{\nu^2}x\right\},
\]
where $\quad \xi=-2\frac{R_0-R_1R_2}{\nu^{2}}-1$. If $R_0=0$ and $2R_1R_2\le \nu^2$, then $\sigma_t\to 0$ a.s.\ as $t\to\infty$.
\end{proposition}
The steady-state distribution therefore belongs to the class of Generalized Inverse Gaussian distributions, see e.g.\ \cite{jorgensen1982statistical}. The integration constant such that $\pi$ integrates to one is provided in the proof (see Section \ref{proof:steady_state}). We distinguish four different cases, based on the values for $R_0$ and $R_1$:
\begin{enumerate}
\item For $R_0=R_1=0$, the process $\sigma_t$ becomes a geometric Brownian motion without drift, which goes to zero almost surely as $t\to\infty$.
\item  For $R_0>0$ and $R_1=0$, the process $\sigma_t$ becomes a linear diffusion and we recover the inverse gamma distribution as steady-state distribution, see e.g.\ \cite{barone2005option}. The first moment of $\pi$ is equal to $R_2$, so that $R_2$ can be interpreted as the long-term level of mean reversion. Higher order moments do not always exist because the inverse gamma distribution has a right tail with polynomial decay. Remark that $\pi(0)=0$, regardless of $R_0$ and $R_2$, due to the exponential decay of of the left tail.

\item For $R_0=0$ and $R_1>0$, the process $\sigma_t$ becomes a logistic diffusion. If $2R_1R_2\le \nu^2$, then a similar behavior as in the first case occurs and $\sigma_t\to 0$ almost surely as $t\to \infty$. If $2R_1R_2>\nu^2$, we recover the gamma distribution as steady-state distribution, which has finite moments of any order. In particular, the first moment equals $R_2-\frac{\nu^2}{2R_1}$. As highlighted in \cite{merton1975asymptotic} and \cite{ewald2007geometric}, we can therefore no longer interpret $R_2$ as the long-term level of mean reversion. Remark that $\pi(0)=0$ if and only if $R_1R_2>\nu^2$.

\item For $R_0,R_1>0$, the steady-state distribution has a gamma tail on the right and an inverse gamma tail on the left. As a consequence, $\pi$ has finite moments of any order and $\pi(0)=0$. In particular, the first moment equals (see e.g., \cite{jorgensen1982statistical})
\[
\frac{\sqrt{R_0R_2}}{\sqrt{R_1}}\frac{K_{\xi+1}(4\sqrt{R_0R_1R_2}\nu^{-2})}{K_\xi(4\sqrt{R_0R_1R_2}\nu^{-2})},
\]
where $K_\xi$ denotes the modified Bessel function of the second kind.\footnote{In Appendix \ref{appendix:lower_bound_first_moment} we provide a tight lower bound (based on Jensen's inequality) for the first moment of the steady-state distribution that does not involve any special functions.} In general, the first moment of the steady-state distribution will not be exactly equal to $R_2$, so we can not interpret $R_2$ as the long-term level of mean-reversion. However, the difference is small for standard parameterizations.
\end{enumerate}

\section{Moment explosions} \label{sec:moment_explosions}
In this section, we build on the general findings of \cite{lions2007correlations}  to investigate moment explosions of the stock price in our model. In order to use \eqref{eq:SDE_x}-\eqref{eq:SDE_v} for option pricing, we need $S_t$ to be a $\Q$-martingale. The following proposition derives a necessary and sufficient condition.
\begin{proposition} \label{prop:true_martingale}
$S_t$ is a $\Q$-martingale if and only if $R_1 \ge \rho\nu$.
\end{proposition}
If $R_1=0$, then $S_t$ is a $\Q$-martingale if and only if $\rho\le 0$, which is a well known problem with this type of model. While equity markets generally feature a negative correlation between stock returns and volatility, other applications might require a positive correlation. Proposition \ref{prop:true_martingale} shows that our model can accommodate a positive correlation if $R_1$ is sufficiently large. In particular, if $R_1\ge \nu$, then $S_t$ is always a $\Q$-martingale in our model, regardless of $\rho$. Intuitively, the quadratic drift term has a stabilizing effect on the volatility because the speed of mean reversion becomes very large at high volatility levels.

\cite{andersen2007moment} highlight the importance for $S_t$ to have finite moments greater than one for pricing contracts with super-linear payoff, which occur frequently in interest rate derivatives markets. Examples include CMS swaps, in-arrears swaps, and Eurodollar futures. Moreover, when using Monte-Carlo simulations to find the price of a derivative, the payoff needs to have finite second order moment in order to derive confidence intervals on the Monte-Carlo estimator with the central limit theorem. The following proposition derives a lower bound on $R_1$ such that $S_t$ has finite moments of a given order.

\begin{proposition}\label{prop:finite_moments}
Let $m\in\R\setminus [0,1]$. 
\begin{enumerate}
\item If $R_1>\nu(\rho m +\sqrt{m^2-m})$, then 
\[
\E_t\left[S_T^m\right]<\infty,\quad \forall T>t.
\]
If $R_0\ge R_1R_2$, then the statement is also true for $R_1=\nu(\rho m +\sqrt{m^2-m})$.
\item If $R_1<\nu(\rho m +\sqrt{m^2-m})$, then 
\[
\E_t\left[S_T^m\right]=\infty,\quad \forall T>t.
\]
\end{enumerate}
\end{proposition}
In particular, if $R_1=0$ and $m>1$, then $\E_t[S_T^m]$ is finite if and only if $\rho\le -\sqrt{\frac{m-1}{m}}$. A negative correlation has a dampening effect on the moments of the return process, however it must be sufficiently negative in this case for higher moments to exist. For instance, already for $m=2$ we require $\rho\le -70.71\%$ in case $R_1=0$, which can be quite restrictive. Proposition \ref{prop:finite_moments} shows that the quadratic term in the drift of $\sigma_t$ can take over the role of the negative correlation to stabilize the moments of the return process, which allows the correlation to remain a free parameter. Remark also that for $R_1=0$ and $m<0$, we have $\E_t[S_T^m]=\infty$, regardless of $\rho$. 

The seminal work of \cite{lee2004moment} relates moment explosions to the asymptotic behaviour of the Black-Scholes implied volatility smile as a function of log-moneyness. Specifically, define the critical moments
\begin{equation}
    m_+(T)=\sup\{m\colon S_T^m<\infty\},\quad m_-(T)=\inf\{m\colon S_T^m<\infty\}.
    \label{eq:critical_moments}
\end{equation}
Remark that Proposition \ref{prop:finite_moments} implies in particular that the critical moments in our model do not depend on the time horizon $T$, so henceforth we omit the time argument and simply write $m_{\pm}$. Let $\sigma_{BS}(T,x)$ denote the Black-Scholes implied volatility of a European call option with time-to-maturity $T$ and strike price $S_0\e^x$. Using the formulation of \cite{keller2011moment}, the critical moment formula of \cite{lee2004moment} states
\begin{align}
    \limsup_{x\to-\infty} \frac{\sigma^2_{BS}(T,x)}{\lvert x \rvert}=\frac{\beta(-m_-)}{T}\quad \text{and}\quad
    \limsup_{x\to \infty} \frac{\sigma^2_{BS}(T,x)}{\lvert x \rvert}=\frac{\beta(m_+-1)}{T},
    \label{eq:lee_moment_formula}
\end{align}
where we define the decreasing function $\beta\colon \R_+ \to [0,2], x\mapsto 2-4(\sqrt{x^2+x}-x)$. The critical moments in our model can directly be computed using the result of Proposition \ref{prop:finite_moments}, as shown in the following corollary. Note that Black-Scholes implied volatility only makes sense if $S_t$ is a $\Q$-martingale, so we only consider the case $R_1\ge \rho\nu$, cf.\ Proposition \ref{prop:true_martingale}.
\begin{corollary}\label{prop:critical_moments}
Suppose $S_t$ is a $\Q$-martingale, i.e., $R_1\ge \rho\nu$.
\begin{enumerate}
    \item If $|\rho|<1$, then
    \begin{equation}
        m_{\pm}=\frac{1-2\frac{R_1}{\nu}\rho \pm \sqrt{(1-2\frac{R_1}{\nu}\rho)^2+4(1-\rho^2)\frac{R_1^2}{\nu^2}}}{2(1-\rho^2)}.
        \label{eq:critical_moments_model}
    \end{equation}
    \item If $\rho=1$, then $m_-=-\infty$ and $m_+=\frac{R_1^2}{2R_1\nu-\nu^2}$.
    \item If $\rho=-1$, then $m_+=\infty$ and $m_-=\frac{R_1^2}{-2R_1\nu-\nu^2}$.
\end{enumerate}
\end{corollary}
The critical moment formula \eqref{eq:lee_moment_formula} and Corollary \ref{prop:critical_moments} give us important information about the tail behaviour of $x\mapsto \sigma^2_{BS}(T,x)T$ in our model. If $|\rho|<1$, then the critical moments are finite, which implies asymptotically linear behaviour of $\sigma_{BS}^2(T,x)$ in $x$ for all $T>0$. The slope of the small and large strike tail is controlled by both $\rho$ and $\frac{R_1}{\nu}$. For $R_1=0$, we get in particular $m_+=(1-\rho^2)^{-1}$ and $m_-=0$. In this case, $\rho$ only controls the slope of large strike tail, while the slope of the small strike tail is always equal to $\beta(0)=2$. With $R_1$ as a free parameter, we can therefore more accurately capture both the small and the large strike tail of the Black-Scholes implied volatility skew. 

\begin{remark}
As noted by \cite{lee2004moment}, the critical moment formula \eqref{eq:lee_moment_formula} can be useful to facilitate model calibration. Suppose we observe a Black-Scholes implied volatility skew for a range of strikes and a certain maturity $T>0$. From the smallest and largest strike, we can approximately infer $m_-$ and $m_+$, respectively.\footnote{Remark that in our model, the critical moments do not depend on the time horizon, while the implied critical moments will likely not be exactly equal for different option maturities, in which case we can for example average the implied critical moments across maturities.} The parameters $\rho$ and $\frac{R_1}{\nu}$ can then be calibrated to these implied critical moments using \eqref{eq:critical_moments_model}. This approach should be seen as a way to get good initial guesses for $\rho$ and $\frac{R_1}{\nu}$. 
\end{remark}

We end this section with an additional result on the two extreme correlation cases.
\begin{proposition}\label{prop:bounded_price}
\leavevmode
\begin{enumerate}
\item 
If $\rho=-1$ and $R_0\ge R_1R_2$, then 
\[
S_T\le S_t \exp\left\{\frac{\sigma_t}{\nu}+\frac{R_0R_2}{\nu}(T-t)\right\},\quad \forall T> t.
\]
\item
If $\rho=1$, $R_0\ge R_1R_2$, and $2 R_1\ge \nu$, then 
\[
S_T\ge S_t \exp\left\{-\frac{\sigma_t}{\nu}-\frac{R_0R_2}{\nu}(T-t)\right\},\quad \forall T> t.
\]
\end{enumerate}
\end{proposition}
For $\rho=-1$, we know from Proposition \ref{prop:true_martingale} that $S_t$ is a $\Q$-martingale and from Proposition \ref{prop:finite_moments} that $\E_t[S_T^m]<\infty$ for all $m>1$ and all $T>t$. If moreover $R_0\ge R_1R_2$, then Proposition \ref{prop:bounded_price} shows that the stock price becomes bounded form above. Remark that this additional condition is trivially satisfied when $R_1=0$. For $\rho=1$, the stock price is a $\Q$-martingale if and only if $R_1\ge \nu$, see Proposition \ref{prop:true_martingale}. From Proposition \ref{prop:finite_moments} we have in this case $\E_t[S_T^m]<\infty$ for all $m<0$ and all $T>t$. If moreover $R_0\ge R_1R_2$, then the stock price has a lower bound strictly larger than zero. Remark that these results are consistent with the critical moments derived in Corollary \ref{prop:critical_moments}.

\section{A polynomial diffusion in disguise} \label{sec:polynomial_disguise}
In this section we show how our model can be related to the class of polynomial diffusions using a conveniently chosen change of measure.

Define the process $y_t$ through the following stochastic differential equation (SDE)
\begin{equation}
\dd y_t = -\frac{1}{2}z^2 \sigma_t^2\,\dd t +z\sigma_t\,\dd W_t,\quad y_0=0,
\end{equation}
with $z=\frac{R_1}{\nu}$. Fix a time horizon $T>0$ and define the probability measure $\Q^z$ through the following Radon-Nikodym derivative
\begin{equation}
\frac{\dd \Q^z}{\dd \Q}=\e^{y_T}=\e^{ -\frac{1}{2}z^2\int_0^T\sigma_t^2\,\dd t+z\int_0^T\sigma_t\dd W_t}.
\end{equation} 
Remark $\Q^z=\Q$ if $R_1=0$. The following proposition shows that the change of measure is well defined.
\begin{proposition}\label{prop:RN_integrability}
The process $\e^{y_t}$ is a $\Q$-martingale.
\end{proposition}

Henceforth $\E^z_t[\cdot]$ denotes the $\Fcal_t$-conditional $\Q^z$-expectation. By Girsanov's theorem we have that 
\[
W_t^z=W_t-z\int_0^t \sigma_s\,\dd s\quad \text{and}\quad B_t^z=B_t
\]
are independent $\Q^z$-Brownian motions. The $\Q^z$-dynamics of $\sigma_t$ becomes
\begin{equation}
\dd \sigma_t = (R_0R_2+\sigma_t(R_1R_2-R_0))\,\dd t+\nu\sigma_t\,\dd W_t^z.
\end{equation}
The quadratic term in the drift of $\sigma_t$ vanishes and $\sigma_t$ becomes a polynomial diffusion under $\Q^z$. Indeed, it has an affine drift and a linear dispersion function, so that its infinitesimal generator maps polynomials to polynomials of the same degree or less. This allows us to compute all $\Q^z$-moments of $\sigma_t$ in closed form, which is informative about the $\Q^z$-distribution of $\sigma_t$. For derivative pricing purposes (see Section \ref{sec:option_pricing} for more details), we are not particularly interested in the $\Q^z$-distribution of $\sigma_t$. Instead, we are mainly interested in the $\Q$-distribution of $x_t$ or, equivalently, in the joint $\Q^z$-distribution of $x_t$ and $y_t$. The process $(x_t,y_t,\sigma_t)$ is not a polynomial diffusion under $\Q^z$, because the drift of $x_t$ and $y_t$ contains a quadratic term $\sigma_t^2$:
\begin{align*}
&\dd x_t = (z\rho-\frac{1}{2})\sigma_t^2\,\dd t +\sigma_t (\rho\,\dd W_t^z+\sqrt{1-\rho^2}\,\dd B_t^z),\\
&\dd y_t = \frac{1}{2}z^2\sigma_t^2\,\dd t +z \sigma_t\,\dd W_t^z,\\
&\dd \sigma_t = (R_0R_2+\sigma_t(R_1R_2-R_0))\,\dd t+\nu\sigma_t\,\dd W_t^z.
\end{align*}
However, by augmenting the state with $\sigma_t^2$, we can see that $(x_t,y_t,\sigma_t,\sigma_t^2)$ jointly becomes a polynomial diffusion under $\Q^z$ since $\sigma_t^2$ has the following dynamics
\[
\dd \sigma_t^2 = (2R_0R_2\sigma_t+\sigma_t^2(2R_1R_2-2R_0+\nu^2))\,\dd t+2\nu\sigma_t^2\,\dd W_t^z.
\]
This observation makes it possible to calculate all conditional $\Q^z$-moments of $(x_t,y_t,\sigma_t)$ in closed form. Before we do this, we first introduce some notation. Denote for $m,n\in\N$ by $\mathrm{Pol}_m(\R^n)$ the space of polynomials on $\R^n$ with total degree at most $m$. Define the subspace $P_{m}\subset \mathrm{Pol_{2m}(\R^3)}$ of trivariate polynomials as
\begin{align*}
P_{m}=\left\{ (x,y,z)\mapsto p(x,y)q(z)\;\vert \; p\in \mathrm{Pol}_m(\R^2),q\in \mathrm{Pol}_{2(m-\mathrm{deg}(p))}(\R)\right\},
\end{align*}
where $\mathrm{deg}(\cdot)$ denotes the total degree of a polynomial.  The following lemma provides the dimension of $P_m$, i.e.\ the number of linearly independent polynomials in $P_m$.

\begin{lemma}\label{lemma:dimension}
The dimension of $P_m$ is
\[d_m=\dim(P_m)=\frac{1}{3}m^3+\frac{3}{2}m^2+\frac{13}{6}m+1.\]
\end{lemma}

The following proposition provides an explicit formula for the conditional $\Q^z$-moments of $(x_t,y_t,\sigma_t)$, which will be the cornerstone of the derivative pricing approximation method in Section \ref{sec:option_pricing}.
\begin{proposition}\label{prop:moments}
The infinitesimal generator $\Gcal$ of the process $(x_t,y_t,\sigma_t)$ under $\Q^z$ leaves $P_m$ invariant. That is, there exists a matrix $G_m\in \R^{d_m\times d_m}$, such that $\Gcal H_m= G_m H_m$, where $H_m=(h_1,\ldots,h_{d_m})^\top$ denotes a vector of polynomial basis functions for $P_m$. As a consequence, we have for any $t\le T$
\begin{equation}
\E_t^z\left[H_m\left(x_T,y_T,\sigma_T\right)\right]=\e^{G_m(T-t)}H_m\left(x_t,y_t,\sigma_t\right).\label{eq:moment_formula}
\end{equation}
\end{proposition}
The matrix $G_m$ is straightforward to construct in practice by choosing $H_m$ to be a monomial basis and then collecting terms according to their exponents in the vector of polynomials $\Gcal H_m$, see equation \eqref{eq:generator} in the Appendix.


\section{Derivative pricing}\label{sec:option_pricing}
In this section we show how European style derivatives on the stock price $\e^{x_T}$ can efficiently be computed using the available $\Q^z$-moments of $(x_T,y_T)$.
\subsection{Polynomial payoff approximation}
Consider a derivative on the stock price with payoff $F(\e^{x_T})$ at time $T>0$, for some integrable payoff function $F$. The price at time $0$ is given by
\begin{equation}
\pi=\E_0[F(\e^{x_T})]=\E_0^z[\e^{-y_T}F(\e^{x_T})].
\label{eq:derivative_price}
\end{equation}
The auxiliary process $y_t$ can therefore be interpreted as a stochastic discount rate under the new measure. The positive correlation between $y_t$ and $\sigma_t$ provides a dampening effect on the `discounted' payoff under the new measure, which is the equivalent of the dampening effect of the quadratic drift term of $\sigma_t$ that disappeared with the measure change. 

The conditional $\Q$-distribution of $x_T$ is not known, but we do know all the conditional $\Q^z$-moments of $(x_T,y_T)$ thanks to the moment formula \eqref{eq:moment_formula}. Therefore, we can approximate the derivative price by approximating the function $(x,y)\mapsto e^{-y}F(\e^x)$ with a polynomial $p_n \in \mathrm{Pol}_n(\R^2)$, for some $n\in\N$. We would like the polynomial approximation to be most accurate for the values that $(x_T,y_T)$ is most likely to take under $\Q^z$, since they contribute the most to the right hand side of \eqref{eq:derivative_price}. This motivates the following least-squares approach to determine the approximating polynomial
\begin{equation}
p_n=\underset{p\in\mathrm{Pol}_n(\R^2)}{\mathrm{arg\,min}} \int_{\R^2} (e^{-y}F(e^x)-p(x,y))^2w(x,y)\,\dd x\,\dd y,
\label{eq:optimal_pol}
\end{equation}
where $w$ is an auxiliary probability density function which proxies the unknown $\Q^z$-density of $(x_T,y_T)$.\footnote{We assume that $w$ is such that the double integral in \eqref{eq:optimal_pol} is finite for all $p\in\mathrm{Pol}_n(\R^2)$.} Put differently, $p_n$ is the orthogonal projection of $(x,y)\mapsto e^{-y}F(\e^x)$ on the space of bivariate polynomials of total degree $n$ or less in a weighted Hilbert function space with weight $w$. If we denote by $\vec{p}_n\in\R^{d_n}$ the vector representation of $p_n$ with respect to the basis $H_n$, the option price approximation becomes
\begin{equation}
\pi\approx \pi_n=\vec{p}_n^\top \e^{G_n T}H_n(x_0,y_0,\sigma_0).
\label{eq:approx_series}
\end{equation}
In Section \ref{sec:aux_dens} we show how to choose $w$ and in Section \ref{sec:optimal_pol} we solve the optimization problem in \eqref{eq:optimal_pol}.

\subsection{The auxiliary density}\label{sec:aux_dens}
It remains to choose a good auxiliary density $w$. We use an approach that closely resembles the Gaussian mixture specification of \cite{ackerer2019option}. Conditional on the trajectory $\{W_t^z,t\le T\}$, the $\Q^z$-density function of the random variable $(x_T,y_T)$ can be formally written as
\[
(x,y)\mapsto \phi_{M_T,V_T}(x)\delta(y-y_T),
\]
where $\phi_{M_T,V_T}$ denotes the density function of a Gaussian distribution with mean $M_T$ and variance $V_T$, $\delta$ denotes the Dirac delta function, and 
\begin{align*}
M_T=x_0+(-\frac{1}{2}+z\rho)\int_0^T\sigma_s^2\,\dd s+\rho\int_0^T\sigma_s\,\dd W_s^z,\quad
V_T=(1-\rho^2)\int_0^T\sigma_s^2\,\dd s
\end{align*}
The true $\Q^z$-density function of $(x_T,y_T)$ can therefore be expressed as
\begin{equation*}
(x,y)\mapsto\E^z_0[\phi_{M_T,V_T}(x)\delta(y-y_T)].
\end{equation*}
We specify the auxiliary density as
\begin{equation}
w(x,y)=\sum_{k=1}^{\Kcal} w^{(k)} \phi_{m^{(k)},v^{(k)}}(x)\delta(y-y^{(k)}),\label{eq:aux_dens}
\end{equation}
where $m^{(k)},y^{(k)}\in\R$, $v^{(k)}\in\R_+$, $w^{(k)}\in[0,1]$, $k=1,\ldots,\Kcal$, are constants to be determined subject to $\sum_{k=1}^{\Kcal} w^{(k)}=1$. The quadruplets $(w^{(k)},m^{(k)},v^{(k)},y^{(k)})$ represent a discretization of the $\Q^z$-distribution of $(M_T,V_T,y_T)$ in $\Kcal\ge 1$ mass points, which can be obtained by discretizing the single source of uncertainty $\{W_t^z,t\le T\}$. Specifically, we use the IJK scheme of \cite{kahl2006fast} with $d\ge 1$ equidistant time steps to obtain the following discretization scheme for $\sigma_t$, $M_t$, $V_t$, and $y_t$:
\begin{align*}
    &\hat{\sigma}_{{n+1}}=\hat{\sigma}_{n}+(R_0R_2+\hat{\sigma}_{n}(R_1R_2-R_0))\Delta + \nu \hat{\sigma}_{n}\sqrt{\Delta} Z_{n+1}+\frac{1}{2}\nu^2\hat{\sigma}_{n}(\Delta Z_{n+1}^2-\Delta),\\
    &\hat{M}_{{n+1}}=\hat{M}_{{n+1}}+(-\frac{1}{2}+z\rho)\frac{\hat{\sigma}_{{n+1}}^2+\hat{\sigma}_n^2}{2}\Delta+\rho \hat{\sigma}_{t_{n}}\sqrt{\Delta}Z_{n+1},\\
    &\hat{V}_{n+1}=\hat{V}_n+(1-\rho^2)\frac{\hat{\sigma}_{{n+1}}^2+\hat{\sigma}_n^2}{2}\Delta, \\
    &\hat{y}_{n+1}=\hat{y}_n+\frac{1}{2}z^2\frac{\hat{\sigma}_{{n+1}}^2+\hat{\sigma}_n^2}{2}\Delta+z\hat{\sigma}_{n}\sqrt{\Delta}Z_{n+1},
\end{align*}
where $\Delta =\frac{T}{d}$ is the step size, $(Z_1,\ldots,Z_d)$ is a $d$-dimensional standard normal random variable, and $\hat{\sigma}_0=\sigma_0$, $\hat{M}_0=M_0$, $\hat{V}_0=V_0$, $\hat{y}_0=y_0$.
If we are given $\Kcal$ weighted samples $(w^{(k)},Z_1^{(k)},\ldots,Z_d^{(k)})$, $k=1\ldots,\Kcal$, of the random variable $(Z_1,\ldots,Z_d)$, then by plugging them into the above scheme we obtain the quadruplets $(w^{(k)},m^{(k)},v^{(k)},y^{(k)})$. As highlighted by \cite{ackerer2019option}, raw Monte-Carlo simulation with $w^{(k)}\equiv 1/\Kcal$ requires far too many samples to produce an accurate approximation of the distribution. Instead, deterministic discretizations of the $d$-dimensional standard normal distribution, such as the quantization techniques of \cite{pages2003optimal} or Gaussian cubature rules, are preferred in order to keep $\Kcal$ small. In the numerical study in Section \ref{sec:numerical}, we use the multivariate Gauss-Hermite quadrature method described in \cite{jackel2005note} to obtain the weighted samples $(w^{(k)},Z_1^{(k)},\ldots,Z_d^{(k)})$. The advantage of Gauss-Hermite quadrature is that the tails of the distribution are accurately captured, which is important for the stability of our approximation method as $n$, the total polynomial degree of the approximation, increases.

\subsection{The optimal polynomial}\label{sec:optimal_pol}
Now that we have specified the auxiliary density function, we can solve the optimization problem in \eqref{eq:optimal_pol}. Denote by $B_n=(b_1,\ldots,b_{N_n})^\top$, $N_n={n+2\choose 2}$, a vector of polynomial basis functions for $\mathrm{Pol}_n(\R^2)$. We can rewrite \eqref{eq:optimal_pol} as
\begin{equation}
c_n=\underset{c\in\R^{N_n}}{\mathrm{arg\,min}} \int_{\R^2} (e^{-y}F(e^x)-c^\top B_n(x,y))^2w(x,y)\,\dd x\,\dd y.
\label{eq:optimal_pol2}
\end{equation}
\begin{proposition}\label{prop:optimization}
The unique solution of \eqref{eq:optimal_pol2} is $c_n=D^{-1}f$, with
\begin{equation}
D_{i,j}=\int_{\R^2} b_i(x,y)b_j(x,y)w(x,y)\,\dd x\,\dd y,\quad f_i=\int_{\R^2} e^{-y}F(\e^{x})b_i(x,y)w(x,y)\,\dd x\,\dd y,
\label{eq:unique_sol}
\end{equation}
for $i,j=1,\ldots,N_n$.
\end{proposition}

Without loss of generality, we assume that $B_n$ is a monomial basis with $b_i(x,y)=x^{\alpha_i}y^{\beta_i}$, for exponents $\alpha_i,\beta_i\in\N$ such that $\alpha_i+\beta_i\le n$, $i=1,\ldots,N_n$. Plugging \eqref{eq:aux_dens} in the expression for $D_{i,j}$ in \eqref{eq:unique_sol} gives
\[
D_{i,j}=\sum_{k=1}^K w^{(k)}(y^{(k)})^{\beta_i+\beta_j}\int_{\R}x^{\alpha_i+\alpha_j}\phi_{m^{(k)},v^{(k)}}(x)\,\dd x.
\]
The remaining integral is simply the $(\alpha_i+\alpha_j)$-th moment of the univivariate Gaussian distribution, which is known in closed form.\footnote{In Appendix \ref{appendix:moments_gaussian} we provide a simple formula for the moments of the Gaussian distribution.} The elements of the vector $f$ become
\begin{equation}
f_i=\sum_{k=1}^K w^{(k)}\e^{-y^{(k)}}(y^{(k)})^{\beta_i}\int_{\R} F(\e^{x}) x^{\alpha_i}\phi_{m^{(k)},v^{(k)}}(x)\,\dd x.
\label{eq:payoff_coeff}
\end{equation}
In general, the integral in \eqref{eq:payoff_coeff} has to be computed numerically, for example using Gauss-Hermite quadrature. For specific payoff functions, the integral can be computed in closed form. For example, the following proposition derives a recursive formula for the case of a European call option.
\begin{proposition}\label{prop:recursion}
Suppose $F(x)=(\e^{x}-K)^+$, for some $K>0$. The integral
\[
I_{n}^{(k)}=\int_{\R} F(\e^{x}) x^n\phi_{m^{(k)},v^{(k)}}(x)\,\dd x,
\]
satisfies the following recursion for $n\ge 1$
\begin{align*}
I_{n}^{(k)}&=(m^{(k)}+v^{(k)})I_{n-1}^{(k)}+v^{(k)}(n-1)I_{n-2}^{(k)}+Kv^{(k)} J_{n-1}^{(k)},\\
J_{n}^{(k)}&=m^{(k)}J_{n-1}^{(k)}+v^{(k)}(n-1)J_{n-2}^{(k)}+\sqrt{v^{(k)}}(\log(K))^{n-1}\phi(\xi^{(k)}),
\end{align*}
with $\xi^{(k)}=\frac{m^{(k)}-\log (K)}{\sqrt{v^{(k)}}}$, $\phi$ the standard normal density, and starting values
\begin{align*}
I_{-1}^{(k)}=0, \quad I_{0}^{(k)}=\e^{m^{(k)}+\frac{1}{2}v^{(k)}}\Phi\left(\xi^{(k)}+\sqrt{v^{(k)}}\right)-K\Phi(\xi^{(k)}),\quad 
J_{-1}^{(k)}=0,\quad J_0^{(k)}=\Phi(\xi^{(k)}),
\end{align*}
with $\Phi$ the standard normal cumulative distribution function.
\end{proposition}

\section{Numerical study}\label{sec:numerical}
In this section we investigate the numerical accuracy of the option price approximation proposed in the previous section.

\begin{figure}
    \centering
    \includegraphics[width=0.7\textwidth]{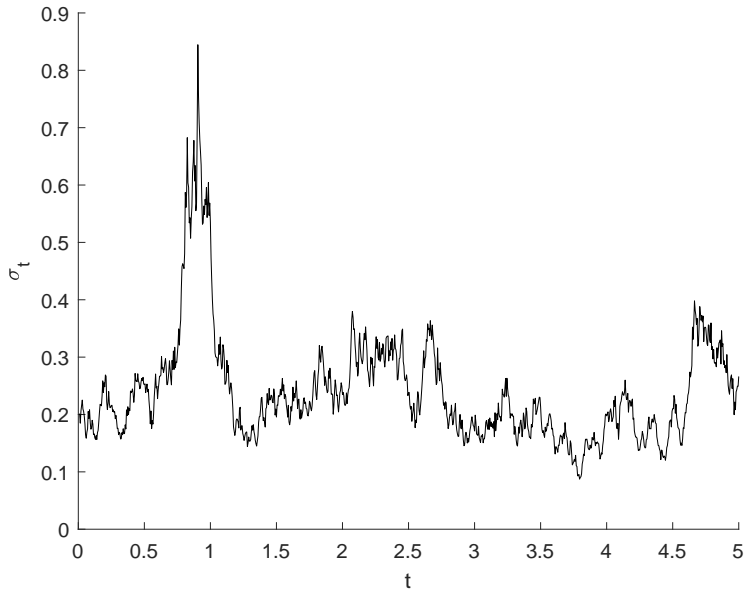}
    \caption{Simulated trajectory for $\sigma_t$ with parameters $R_0=R_1=5$, $R_2=\sigma_0=0.20$, $\nu=1$.}
    \label{fig:vol_trajectory}
\end{figure}

\begin{figure}
    \centering
    \begin{subfigure}[b]{0.45\textwidth}
        \includegraphics[width=\textwidth]{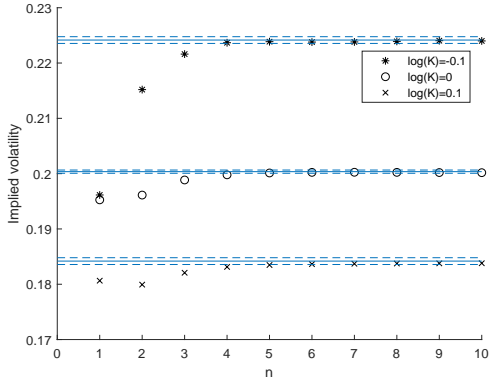}
        \caption{$T=1/12$, $d=1$}
        \label{fig:mom_conv_1m_d1}
    \end{subfigure}
    \begin{subfigure}[b]{0.45\textwidth}
        \includegraphics[width=\textwidth]{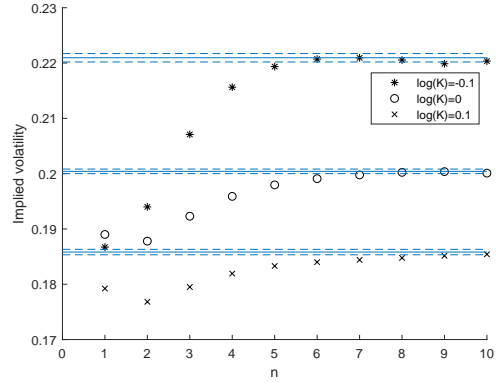}
        \caption{$T=2/12$, $d=1$}
        \label{fig:mom_conv_2m_d1}
    \end{subfigure}    
    \begin{subfigure}[b]{0.45\textwidth}
        \includegraphics[width=\textwidth]{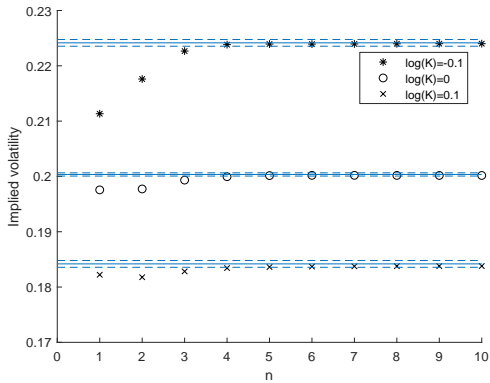}
        \caption{$T=1/12$, $d=2$}
        \label{fig:mom_conv_1m_d2}
    \end{subfigure}
    \begin{subfigure}[b]{0.45\textwidth}
        \includegraphics[width=\textwidth]{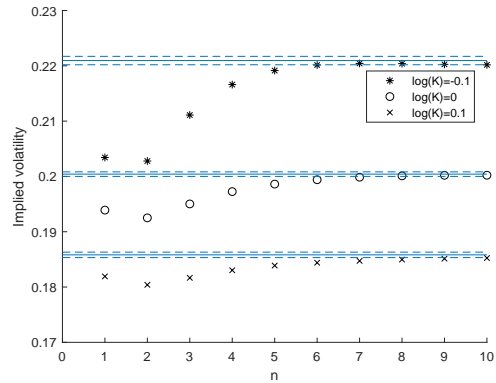}
        \caption{$T=2/12$, $d=2$}
        \label{fig:mom_conv_2m_d2}
    \end{subfigure}
    \caption{Black-Scholes implied volatilities of approximated European call option prices with time-to-maturity of one and two months for varying number of terms $n$ in the series. Solid blue lines are Monte-Carlo estimators using $10^6$ sample paths and the dashed blue lines are the corresponding 99\% confidence intervals. We use a quadratic polynomial approximation of the discounted payoff as a control variate to substantially reduce the variance of the Monte-Carlo estimator. The top row uses a single time step discretization to construct the auxiliary distribution, while the bottom row uses two equidistant time steps.}
    \label{fig:mom_conv}
\end{figure}

We set the model parameters as $R_0=R_1=5$, $\nu=1$, $R_2=\sigma_0=0.20$, $\rho=-0.5$, $x_0=0$. These are realistic parameters that produce a volatility process with strong mean-reversion and a high volatility of volatility that can cause occasional spikes, see for example Figure \ref{fig:vol_trajectory} for a simulated (under $\Q$) trajectory. Consider a European call option with time-to-maturity $T\in\{1/12,2/12\}$ and log-strike $\log(K)\in\{-0.1,0,0.1\}$.  Figure \ref{fig:mom_conv_1m_d1} and \ref{fig:mom_conv_2m_d1} plot the option price approximations $\pi_n$ for $n$ ranging from 1 to 10. We set $d=1$, and use the Gauss-Hermite quadrature rule to obtain a discretization of the univariate standard normal distribution in $\Kcal=15$ points. As a benchmark, we also run a Monte-Carlo simulation with $10^6$ sample paths.\footnote{We use a quadratic polynomial approximation of the discounted payoff as a control variate to substantially reduce the variance of the Monte-Carlo estimator. To determine the polynomial approximation, we perform a linear regression with the simulated trajectories. This is similar to the polynomial approximation in Section \ref{sec:option_pricing}, where we now use the simulated empirical distribution as auxiliary distribution.} For all strikes and maturities considered, $\pi_n$ converges to within the confidence bands of the Monte-Carlo estimator with $n\le 10$. For $n<3$, the pricing error is most noticeable for the low strike option (i.e., the in-the-money call). This is not surprising, since the true log-return distribution is negatively skewed ($\rho<0$). Therefore, approximations which do not take into account at least third order moments will be far off for low strike options. The results are robust to changes in the number of discretization points $\Kcal$, as long as it is not too small. If $\Kcal$ is chosen very small (say, $\Kcal=3$), then the approximation blows up for larger $n$. Intuitively, for small $\Kcal$ the auxiliary distribution $w$ has very thin tails and the polynomial approximation of the discounted payoff will therefore only be accurate over a small domain. Since the true probability distribution assigns considerable weight outside of this domain, the polynomial approximation will blow up quickly. In Figure \ref{fig:mom_conv_1m_d2} and \ref{fig:mom_conv_2m_d2}, we do the same exercise with $d=2$. For the auxiliary distribution, we use the Gauss-Hermite quadrature rule with 15 points in each dimension, which gives a total of $\Kcal=15^2=225$ points.\footnote{Using the pruning method described in \cite{jackel2005note}, we can reduce the number of discretization points to 185 by omitting the `corner' points that carry a very small weight.} Compared to the case $d=1$, the approximations converges faster to the true price. However, this comes at a computational cost because the number of discretization points in the auxiliary distribution is much larger.

\section{Conclusion} \label{sec:conclusion}
We have introduced a new stochastic volatility model featuring a volatility process with a quadratic drift and a linear dispersion function. We have shown that the quadratic term in the drift is important to control moment explosions in the stock price and, in particular, the small strike tail of the Black-Scholes implied volatility skew. The volatility process has a stationary distribution that belongs to the class of Generalized Inverse Gaussian distributions, which arises frequently in the empirical literature on volatility modeling. In order to make the model tractable, we introduced a change of measure such that the model fits into the class of polynomial diffusions, which opened the door to polynomial expansion methods to accurately approximate option prices. 
 
\clearpage 
\appendix

\section{Proofs}

\subsection{Proof of Proposition \ref{prop:existence_uniqueness}}
We start by showing that \eqref{eq:SDE_v} has a unique $(0,\infty)$-valued solution. We denote by
\[
\mu(x)=(R_0+R_1 x)(R_2-x)\quad \text{and}\quad \Sigma(x)=\nu x
\]
the drift and dispersion function of $\sigma_t$, respectively. Since $\mu$ and $\Sigma$ are polynomials, they are in particular locally Lipschitz continuous. Hence, strong uniqueness holds for solutions of \eqref{eq:SDE_v}, see e.g.\ \cite[Theorem 2.5]{karatzas1991brownian}. The dispersion function satisfies a linear growth condition \[
|\Sigma(x)|^2\le K_1(1+|x|^2),\]
for $K_1\ge\nu$. The drift function does not satisfy a linear growth condition, so the classical existence result of It\^o  (see e.g., \cite[Theorem 2.9]{karatzas1991brownian}) does not apply. However, since the quadratic term in $\mu$ has a negative coefficient, $\mu$ does satisfy
\[
x\mu(x) \le K_2(1+|x|^2),
\]
for some $K_2\ge 0$. Hence, there exists a unique global solution to \eqref{eq:SDE_v}, cf.\ \cite[Chapter 4.5, p.135]{kloeden1005numerical}. It remains to verify that the solution stays in $(0,\infty)$, which we proof using a comparison theorem. Consider the logistic diffusion
\begin{equation}
\dd X_t=(-R_1X_t^2+(R_1R_2-R_0)X_t)\,\dd t+\,\nu X_t\dd W_t,\quad X_0=\sigma_0.
\label{eq:SDE_verhulst}
\end{equation}
This SDE has a unique solution given by
\[
X_t=\frac{Y_t}{1+R_1 \int_0^t Y_s\,\dd s},\quad \text{with}\quad Y_t=X_0\e^{(R_1R_2-R_0-\frac{1}{2}\nu^2)t+\nu W_t}.
\]
Notice that, since $R_1\ge 0$, we have $X_t>0$ for all $t\ge 0$. Using a comparison theorem \cite[ Chapter VI, Theorem 1.1]{ikeda1989stochastic} and $R_0R_2\ge 0$ gives a.s.\ $\sigma_t\ge X_t>0$ for all $t\ge 0$.

\begin{remark}\label{remark:explosion}
Notice that, if $R_1<0$, then $X_t$ explodes in finite time and therefore $\sigma_t$ as well. The assumption $R_1\ge 0$ is therefore crucial to guarantee existence of a global solution.
\end{remark}

Next, we show that the stochastic integrals in \eqref{eq:SDE_x} are well defined by showing that 
\[
\E_t\left[ \int_t^T\sigma_s^2\,\dd s\right]<\infty,\quad \forall T\ge t.
\]
To this end, consider the SDE
\begin{equation*}
\dd Z_t=(R_0R_2+(R_1R_2-R_0)Z_t)\,\dd t+\nu Z_t\,\dd W_t,\quad Z_0=\sigma_0,
\end{equation*}
which has $
Z_t=Y_t\left(1+R_0R_2\int^t_0Y_s^{-1}\,\dd s\right)$ as unique solution. Using a comparison theorem \cite[Chapter VI, Theorem 1.1]{ikeda1989stochastic} and $R_1\ge 0$ gives a.s.\ $\sigma_t\le Z_t$ for all $t\ge 0$. Since $Z_t$ is a polynomial diffusion, it has finite moments of any order. Therefore we have
\[
\E_t\left[ \int_t^T\sigma_s^2\,\dd s\right]
\le \int_t^T\E_t\left[Z_s^2\right]\,\dd s
<\infty,\quad \forall T\ge t.
\]

From the comparison arguments used in first and second part of this proof, we also obtain the following pathwise bounds on $\sigma_t$
\begin{equation}\label{eq:pathwise_bounds}
X_t \le \sigma_t \le Z_t,\quad a.s.
\end{equation}

\subsection{Proof of Proposition \ref{prop:steady_state}}\label{proof:steady_state}
Using the Fokker-Planck equation we have that $\pi$ must satisfy the following second order linear ODE
\[
\frac{\nu^2}{2}\frac{\dd^2}{\dd x^2}[x^2\pi(x)]=\frac{\dd}{\dd x}[(R_0+R_1x)(R_2-x)\pi(x)].
\]
Making the Ansatz $\pi=x^\gamma \exp\{\alpha\frac{1}{x}+\beta x\}$, for some constants $\alpha$, $\beta$, and $\gamma$, and collecting terms gives
\[
\pi(x)=C x^{\xi-1}\exp\left\{-2\frac{R_0R_2}{\nu^2}\frac{1}{x}-2\frac{R_1}{\nu^2}x\right\},
\]
where $C$ is a constant to be determined such that $\pi$ integrates to one. 

\begin{remark}
The motivation for this Ansatz comes from the special cases $R_1=0$ and $R_0=0$. If $R_1=0$, then $\sigma_t$ is a GARCH diffusion, which is known to have the inverse gamma distribution as steady-state distribution. If $R_0=0$, then $\sigma_t$ is a logistic diffusion, which is known to have the gamma distribution as steady-state distribution if $2R_1R_2>\nu^2$. Therefore, the steady-state distribution of $\sigma_t$ must contain the gamma and inverse gamma distribution as special cases.
\end{remark}
If $R_0,R_1>0$, then $\pi$ is a generalized inverse Gaussian distribution and the normalization constant becomes (see e.g., \cite{jorgensen1982statistical})
\[
C=\frac{\left(\frac{R_1}{R_0R_2}\right)^{\xi/2}}{2K_\xi(4\sqrt{R_0R_1R_2}\nu^{-2})},
\]
where $K_\xi$ denotes the modified Bessel function of the second kind. If $R_0=0$ and $2R_1R_2>\nu^2$, then $\pi$ is a gamma density and $C$ therefore becomes
\[
C=\frac{\left(\frac{2R_1}{\nu^{2}}\right)^\xi}{\Gamma(\xi)},
\]
where $\Gamma$ denotes the gamma function. Note that the condition $2R_1R_2>\nu^2$ is equivalent to $\xi>0$. 

For $R_0=0$ and $2R_1R_2\le \nu^2$, recall from $\eqref{eq:pathwise_bounds}$ that we have almost surely the following upper bound
\begin{align*}
\sigma_t\le Z_t=\sigma_0 \e^{(R_1R_2-\frac{1}{2}\nu^2)t+\nu W_t}.
\end{align*}
If $2R_1R_2\ge \nu^2$, then $\e^{(R_1R_2-\frac{1}{2}\nu^2)t+\nu W_t}\to 0$ a.s.\ for $t\to\infty$ and therefore we have $\sigma_t\to 0$ a.s.\ for $t\to\infty$.

\subsection{Proof of Proposition \ref{prop:true_martingale}}
By Theorem 2.4(i) in \cite{lions2007correlations}, $S_t$ is a $\Q$-martingale if
\begin{align*}
\lim_{x\to \infty} \frac{\rho \nu x^2+(R_0+R_1x)(R_2-x)}{x}<\infty.
\end{align*}
The limit can be rewritten as
\begin{align*}
\lim_{x\to \infty} \frac{\rho \nu x^2+(R_0+R_1x)(R_2-x)}{x}
&= \lim_{x\to \infty} \frac{(\rho \nu -R_1) x^2+(R_1R_2-R_0)x+R_0R_2}{x}\\
&=
\lim_{x\to \infty} (\rho \nu -R_1) x+R_1R_2-R_0.
\end{align*}
Therefore, $S_t$ is a $\Q$-martingale if $R_1\ge \rho\nu$. Indeed, if $R_1>\rho\nu$ the limit is $-\infty$ and if $R_1=\rho\nu$ the limit is $R_1R_2-R_0$.

Conversely, by Theorem 2.4(ii) in \cite{lions2007correlations}, $S_t$ is not a $\Q$-martingale if 
\[
\lim_{x\to \infty} \frac{\rho \nu x^2+(R_0+R_1x)(R_2-x)}{\phi(x)}>0,
\]
for some smooth, positive, and increasing function $\phi$ such that $\int_\epsilon^\infty \frac{1}{\phi(x)}\,\dd x<\infty$, $\epsilon>0$. Choosing $\phi(x)=x^2$ gives
\begin{align*}
\lim_{x\to \infty} \frac{\rho \nu x^2+(R_0+R_1x)(R_2-x)}{x^2}
&=\rho\nu-R_1.
\end{align*}
Therefore, $S_t$ is not a $\Q$-martingale if $R_1<\rho\nu$.

\subsection{Proof of Proposition \ref{prop:finite_moments}}
\begin{enumerate}
\item
By Theorem 2.5 in \cite{lions2007correlations}, we need to show that there exists an $A\ge 0$ such that\footnote{The paper of \cite{lions2007correlations} contains some typos that are relevant for the derivation of this proof. Specifically, in equation (26), the function $\beta$ should be defined as $\beta(x)=m\rho\mu(x)x + b(x)$ instead of $\beta(x)=m\mu(x)x + b(x)$. In equation (28), the $\mu$ in the last term has to be replaced by $m$.}
\begin{equation}\label{eq:suff_cond_finite_moments}
\lim_{x\to\infty} -\frac{1}{2}A^2\nu^2x^2-A[m\rho\nu x^2+(R_0+R_1 x)(R_2-x)]-\frac{m^2-m}{2}x^2>-\infty.
\end{equation}
The limit can be rewritten as
\begin{align}
&\lim_{x\to\infty} -\frac{1}{2}A^2\nu^2+x^2-A[m\rho\nu x^2+(R_0+R_1 x)(R_2-x)]-\frac{m^2-m}{2}x^2\nonumber \\
=&
\lim_{x\to\infty} \left(-\frac{1}{2}A^2\nu^2-A[m\rho\nu -R_1]-\frac{m^2-m}{2}\right)x^2-A(R_1R_2-R_0)x-AR_0R_2. \label{eq:suff_cond_finite_moments2}
\end{align}
Define the parabola $f(u)=-\frac{1}{2}u^2\nu^2-u[m\rho\nu -R_1]-\frac{m^2-m}{2}$. If 
\[
m\rho\nu -R_1<-\nu\sqrt{m^2-m},
\]
then $f$ has two distinct positive roots 
\[
A_\pm=\frac{m\rho\nu-R_1\pm \sqrt{(m\rho\nu-R_1)^2-\nu^2(m^2-m)}}{-\nu^2}.
\]
From \eqref{eq:suff_cond_finite_moments2} it becomes clear that if we pick $A\in (A_-,A_+)$, then \eqref{eq:suff_cond_finite_moments} is satisfied.

If 
\[
m\rho\nu -R_1=-\nu\sqrt{m^2-m},
\]
then $f$ only has a single root
\[
A_0=\frac{m\rho\nu - R_1}{-\nu^2}=\frac{\sqrt{m^2-m}}{\nu}> 0.
\]
For all other values, $f$ will be negative. In other words, any value other than $A_0$ will make \eqref{eq:suff_cond_finite_moments2} equal to $-\infty$. It remains to check what happens to \eqref{eq:suff_cond_finite_moments2} for $A=A_0$
\begin{align*}
\lim_{x\to\infty} f(A_0)x^2-A_0(R_1R_2-R_0)x-A_0 R_0R_2
=\lim_{x\to\infty} -A_0(R_1R_2-R_0)x-A_0 R_0R_2.
\end{align*}
Therefore, the limit will be larger than $-\infty$ if $R_0\ge R_1R_2$.

\item
Follows directly from Theorem 2.6 in \cite{lions2007correlations} and 
\begin{align*}
\lim_{x\to\infty} \frac{\nu x}{x}=\nu,\quad 
\lim_{x\to\infty} \frac{(R_0+R_1x)(R_2-x)}{x^2}=-R_1.
\end{align*}
\end{enumerate}

\subsection{Proof of Corollary \ref{prop:critical_moments}}
Define the function $f(m)=\rho m +\sqrt{m^2-m}$ on $\R\setminus (0,1)$. It is readily verified that $f$ is increasing on $[1,\infty)$ with $f(1)=\rho$ and decreasing on $(-\infty,0]$ with $f(0)=0$. 

If $|\rho|<1$, then standard calculations show $\underset{m\to\pm\infty}{\lim} f(m) =\infty$, so by Proposition \ref{prop:finite_moments} there must be a critical moment both in $(-\infty,0]$ and in $[1,\infty)$. In order to find the critical moments, we have to solve the equation
\begin{equation}
\frac{R_1}{\nu}-\rho m=\sqrt{m^2-m},\quad m\in \R\setminus (0,1). \label{eq:proof_critical_moments}
\end{equation}
Squaring both sides shows that a critical moment $m$ has to satisfy
\[
p(m)=m^2(1-\rho^2)+(2\frac{R_1}{\nu}\rho-1)m-\frac{R_1^2}{\nu^2}=0.
\]
If $|\rho|<1$, then $p$ is a convex parabola with $p(0)=-\frac{R_1^2}{\nu^2}\le 0$ and $p(1)=-\rho^2-\frac{R_1^2}{\nu^2}+2\frac{R_1}{\nu}\rho\le 0$, where the second inequality follows from the assumption that $R_1\ge \nu \rho$. Therefore, $p$ has two real roots
\[
m_\pm=\frac{1-2\frac{R_1}{\nu}\rho \pm \sqrt{(1-2\frac{R_1}{\nu}\rho)^2+4(1-\rho^2)\frac{R_1^2}{\nu^2}}}{2(1-\rho^2)}
\]
with $m_-\le 0$ and $m_+\ge 1$. It is directly verified that $m_\pm$ solves \eqref{eq:proof_critical_moments}.

If $\rho=-1$, then $\underset{m\to\infty}{\lim} f(m)=-\frac{1}{2}<0$. Therefore, $\frac{R_1}{\nu}\ge f(m)$ for all $m\ge 1$, so that $m_+=\infty$. Since $\underset{m\to -\infty}{\lim}  f(m)=\infty$, there will be a critical moment in $(-\infty,0]$ and it is given by the single root of $p$:
\[
m_-=\frac{R_1^2}{-2R_1\nu-\nu^2}\le 0.
\]

Similarly, if $\rho=1$, then $\underset{m\to -\infty}{\lim} f(m)=\frac{1}{2}$. Since we assume $R_1\ge \nu$ in this case, we have in particular $\frac{R_1}{\nu}\ge f(m)$ for all $m\le 0$, so that $m_-=-\infty$. Since $\underset{m\to \infty}{\lim}  f(m)=\infty$, there will be a critical moment in $[1,\infty)$ and it is given by the single root of $p$:
\[
m_+=\frac{R_1^2}{2R_1\nu -\nu^2}\ge 1,
\]
where the inequality follows from the assumption $R_1\ge \rho \nu=\nu$.

\subsection{Proof of Proposition \ref{prop:bounded_price}}
\begin{enumerate}
\item 
Solving \eqref{eq:SDE_x} gives
\begin{align*}
S_T&=S_t\exp\left\{-\int_t^T\sigma_s\,\dd W_s -\frac{1}{2}\int_t^T\sigma_s^2\,\dd s\right\}\\
&=S_t\exp\left\{-\frac{1}{\nu}\left(\sigma_T-\sigma_t-\int_t^T(R_0+R_1\sigma_s)(R_2-\sigma_s)\,\dd s\right) -\frac{1}{2}\int_t^T\sigma_s^2\,\dd s\right\}\\
&=S_t\exp\left\{-\frac{1}{\nu}\left(\sigma_T-\sigma_t-R_0R_2(T-t)-(R_1R_2-R_0)\int_t^T \sigma_s\,\dd s\right) -\left(\frac{1}{2}+\frac{R_1}{\nu}\right)\int_t^T\sigma_s^2\,\dd s\right\}\\
&\le S_t\exp\left\{\frac{\sigma_t}{\nu}+\frac{R_0R_2}{\nu}(T-t)\right\}.
\end{align*}
\item 
Similarly as in the first part, solving \eqref{eq:SDE_x} gives
\begin{align*}
S_T&=S_t\exp\left\{\int_t^T\sigma_s\,\dd W_s -\frac{1}{2}\int_t^T\sigma_s^2\,\dd s\right\}\\
&=S_t\exp\left\{\frac{1}{\nu}\left(\sigma_T-\sigma_t-R_0R_2(T-t)-(R_1R_2-R_0)\int_t^T \sigma_s\,\dd s\right) -\left(\frac{1}{2}-\frac{R_1}{\nu}\right)\int_t^T\sigma_s^2\,\dd s\right\}\\
&\ge S_t\exp\left\{-\frac{\sigma_t}{\nu}-\frac{R_0R_2}{\nu}(T-t)\right\}.
\end{align*}
\end{enumerate}

\subsection{Proof of Proposition \ref{prop:RN_integrability}}
The dynamics of $X_t:=z\sigma_t$ becomes
\begin{align*}
\dd X_t&=(R_0+R_1\sigma_t)(R_2z-X_t)\,\dd t + \nu X_t \,\dd W_t\\
&=(R_0+\nu X_t)(\frac{R_2R_1}{\nu}-X_t)\,\dd t +\nu X_t\,\dd W_t\\
&=b(X_t)\,\dd t+\nu X_t\,\dd W_t,
\end{align*}
where we defined the function $b(x)=(R_0+\nu x)(\frac{R_2R_1}{\nu}-x)$. The dynamics of $\e^{y_t}$ becomes
\[
\dd \e^{y_t}=X_t\e^{y_t}\,\dd W_t.
\]
We conclude by Theorem 2.4(i) in \cite{lions2007correlations} that $\e^{y_t}$ is a $\Q$-martingale, since
\begin{align*}
\underset{x\to +\infty}{\mathrm{lim}} \frac{b(x)+ \nu x^2}{x}
&=R_2R_1-R_0<\infty.
\end{align*}

\subsection{Proof of Lemma \ref{lemma:dimension}}
It is well known that $\dim(\mathrm{Pol}_m(\R^n))={m+n \choose n}$. As a consequence, the dimension of the set of polynomials in $\R^n$ with total degree exactly equal to $k\in\N$ is
\[{k+n \choose n}-{k-1+n \choose n}=k+1,\]
where we define ${n-1 \choose n}=0$.
We therefore get
\begin{align*}
\dim(P_m)&=\sum_{k=0}^m (k+1)(1+2(m-k))\\
&=(2m+1)(m+1)+(2m-1)\sum_{k=0}^m k  -2\sum_{k=0}^m k^2\\
&=(2m+1)(m+1)+\frac{1}{2}m(m+1)(2m-1) -\frac{1}{3}m(m+1)(2m+1)\\
&=\frac{1}{3}m^3+\frac{3}{2}m^2+\frac{13}{6}m+1.
\end{align*}

\subsection{Proof of Proposition \ref{prop:moments}}
Without loss of generality, we can use a monomial basis for $P_m$. A generic element in this basis can be represented as $x^\alpha y^\beta z^\gamma$, with $\alpha,\beta,\gamma\in\N$, $\alpha+\beta\le m$ and $\gamma \le 2(m-\alpha-\beta)$. Applying the $\Q^z$-generator $\Gcal$ of $(x_t,y_t,\sigma_t)$ to this monomial gives
\begin{align}
\Gcal x^\alpha y^\beta z^\gamma =&
\alpha(z\rho -\frac{1}{2})x^{\alpha-1}y^\beta z^{\gamma+2}
+\beta\frac{1}{2}z^2 x^\alpha y^{\beta-1}z^{\gamma+2}+\gamma R_0R_2 x^\alpha y^\beta z^{\gamma-1}
+\gamma(R_1R_2-R_0)x^\alpha y^\beta z^\gamma \nonumber\\
&+\frac{1}{2}\alpha(\alpha-1)x^{\alpha-2}y^\beta z^{\gamma+2}+\frac{1}{2}\beta(\beta-1)z^2 x^\alpha y^{\beta-2}z^{\gamma+2} 
+\frac{1}{2}\gamma(\gamma-1)\nu^2 x^\alpha y^\beta z^\gamma\nonumber \\
&+ \alpha\beta z\rho x^{\alpha-1} y^{\beta-1} z^{\gamma+2} 
+\alpha\gamma \nu\rho x^{\alpha-1}y^{\beta}z^{\gamma+1}
+\beta\gamma z \nu x^\alpha y^{\beta-1}z^{\gamma+1}. \label{eq:generator}
\end{align}
It is readily verified by inspecting each of the above monomials that $\Gcal x^\alpha y^\beta z^\gamma \in P_m$.

\subsection{Proof of Proposition \ref{prop:optimization}}
The optimization problem in \eqref{eq:optimal_pol2} is a convex quadratic programming problem. The first order conditions become
\begin{align*}
2\int_{\R^2} (e^{-y}F(e^x)-c^\top B_n(x,y))b_i(x,y)w(x,y)\,\dd x\,\dd y=0,\quad i=1,\ldots,N_n.
\end{align*}
Re-arranging terms we get
\[
\sum_{j=1}^{N_n} \int_{\R^2} c_j b_j(x,y)b_i(x,y)w(x,y)\,\dd x\,\dd y= \int_{\R^2} e^{-y}F(e^x)b_i(x,y)w(x,y)\,\dd x\,\dd y.
\]
In matrix notation this becomes
\[
Dc=f.
\]
Remark that the matrix $D$ positive-definite by construction and therefore invertible.
\subsection{Proof of Proposition \ref{prop:recursion}}
To lighten the notation, we suppress the superscript $(k)$ throughout the proof. Using the identity $x\phi(x)=-\phi'(x)$ and integrating by parts gives
\begin{align*}
I_{n}&=\int_{\R} (\e^{x}-K)^+ x^n\phi_{m,v}(x)\,\dd x\\
&=\int_{\R} (\e^{m+\sqrt{v}x}-K)^+ (m+\sqrt{v}x)^n\phi(x)\,\dd x\\
&=\int_{-\xi}^\infty (\e^{m+\sqrt{v}x}-K) (m+\sqrt{v}x)^n\phi(x)\,\dd x\\
&=m I_{n-1}+\sqrt{v}\int_{-\xi}^\infty (\e^{m+\sqrt{v}x}-K) (m+\sqrt{v}x)^{n-1}x\phi(x)\,\dd x\\
&=m I_{n-1}-\sqrt{v}\int_{-\xi}^\infty (\e^{m+\sqrt{v}x}-K) (m+\sqrt{v}x)^{n-1}\phi'(x)\,\dd x\\
&=m I_{n-1}-\sqrt{v}\left[-\sqrt{v}(n-1)I_{n-2}-\sqrt{v}\int_{-\xi}^\infty \e^{m+\sqrt{v}x} (m+\sqrt{v}x)^{n-1}\phi'(x)\,\dd x\right]\\
&=(m+v) I_{n-1}+v(n-1)I_{n-2}+Kv\int_{-\xi}^\infty(m+\sqrt{v}x)^{n-1}\phi(x)\,\dd x.
\end{align*}
Define $J_n=\int_{-\xi}^\infty(m+\sqrt{v}x)^{n}\phi(x)\,\dd x$. Similarly as for $I_n$, we use integration by parts to derive the following recursion for $J_n$
\begin{align*}
J_n&=m J_{n-1}-\sqrt{v}\int_{-\xi}^\infty (m+\sqrt{v}x)^{n-1}\phi'(x)\,\dd x\\
&=m J_{n-1}-\sqrt{v}\left[-(m-\sqrt{v}\xi)^{n-1}\phi(-\xi)-(n-1)\sqrt{v}\int_{-\xi}^\infty (m+\sqrt{v}x)^{n-2}\phi'(x)\,\dd x\right]\\
&=mJ_{n-1}+v(n-1)J_{n-2}+\sqrt{v}(\log(K))^{n-1}\phi(\xi).
\end{align*}
For the starting values of the recursion, we have
\begin{align*}
J_0&=\int_{-xi}^\infty \phi(x)\,\dd x=\Phi(\xi),\\
I_0&=\e^{m+\frac{1}{2}v}\Phi\left(\xi+\sqrt{v}\right)-K\Phi(\xi).
\end{align*}
We omit the full derivation of $I_0$ since it is very similar to computing the price of a European call option in the Black-Scholes model.

\section{Auxiliary results}
\subsection{Lower bound on first moment of steady-state density}\label{appendix:lower_bound_first_moment}
Suppose the assumptions of Proposition \ref{prop:steady_state} are satisfied so that a non-trivial steady-state distribution exists, and suppose furthermore that $R_1>0$. We introduce for simplicity the following notation 
\[
\alpha=-2\frac{R_0R_2}{\nu^2},\quad \beta--2\frac{R_1}{\nu^2}.
\]
The first moment of the steady-state density then becomes
\begin{align*}
    \int_0^\infty x\pi(x)\,\dd x&= \frac{\int_0^\infty x^{\xi}\exp\{\alpha \frac{1}{x} +\beta x\}\,\dd x}{\int_0^\infty x^{\xi-1}\exp\{\alpha \frac{1}{x} +\beta x\}\,\dd x}\\
    &=\frac{\int_0^\infty x^{\xi}\exp\{\alpha \frac{1}{x}\}\,\dd\left(\frac{\exp\{\beta x\}}{\beta} \right)}{\int_0^\infty x^{\xi-1}\exp\{\alpha \frac{1}{x} +\beta x\}\,\dd x}\\
    &=-\frac{\xi}{\beta}+\frac{\alpha}{\beta}\frac{\int_0^\infty x^{\xi-2}\exp\{\alpha \frac{1}{x} +\beta x\}\,\dd x}{\int_0^\infty x^{\xi-1}\exp\{\alpha \frac{1}{x} +\beta x\}\,\dd x}\\
    &= -\frac{\xi}{\beta} +\frac{\alpha}{\beta}\int_0^\infty x^{-1} \pi(x)\,\dd x,
\end{align*}
where we have used integration by parts on the integral in the numerator. Remark now that $x\mapsto 1/x$ is strictly convex for $x>0$, so we have by Jensen's inequality
\begin{align*}
     \int_0^\infty x\pi(x)\,\dd x\ge-\frac{\xi}{\beta} +\frac{\alpha}{\beta}\left(\int_0^\infty x\pi(x)\,\dd x\right)^{-1}.
\end{align*}
Remark that the inequality is strict if and only if $R_0\neq 0$. If we denote $\mu=\int_0^\infty x\pi(x)\,\dd x$, we obtain the following quadratic inequality for $\mu$
\[
\mu^2+\frac{\xi}{\beta}\mu-\frac{\alpha}{\beta}\ge 0.
\]
By solving the roots of the parabola, we see that the above inequality can only be true if
\[
\mu\ge \frac{1}{2}\left(-\frac{\xi}{\beta}+\sqrt{\frac{\xi^2}{\beta^2}+4\frac{\alpha}{\beta}}\right).
\]

\subsection{Moments of the Gaussian distribution}\label{appendix:moments_gaussian}
Suppose we want to compute all moments of a univariate Gaussian distribution with mean $\mu\in\R$ and variance $\sigma^2>0$. Consider the Gaussian process $X_t$ defined through the following SDE
\[
\dd X_t=\mu\,\dd t + \sigma\,\dd W_t,\quad X_0=0.
\]
where $W_t$ is a standard Brownian motion. The solution $X_1$ at time 1 has a Gaussian distribution with mean $\mu$ and variance $\sigma^2$. Applying the infinitesimal generator $\Gcal$ of $X_t$ to a monomial $x^n$ gives
\[
\Gcal x^n=n x^{n-1}\mu+\frac{1}{2}n(n-1)x^{n-2}\sigma^2.
\]
Therefore, if we define $G_n\in\R^{(n+1) \times (n+1)}$ as
\[
G_n=
\begin{pmatrix}
0\\
\mu &0\\
\sigma^2&  2\mu & 0\\
0& 3\sigma^2 & 3\mu &0\\
\vdots&\ddots & \ddots & \ddots& \ddots\\
0&\cdots&0& \frac{\sigma^2n(n-1)}{2}& n \mu & 0
\end{pmatrix},
\]
then we can write
\[
\Gcal
\begin{pmatrix}
1 & x & x^2& \cdots & x^n
\end{pmatrix}^\top
=
G_n
\begin{pmatrix}
1 & x & x^2& \cdots & x^n
\end{pmatrix}^\top.
\]
By definition of the generator, we get the following simple formula for the Gaussian moments
\[
\E_0[
\begin{pmatrix}
1 & X_1 & X_1^2& \cdots & X_1^n
\end{pmatrix}
^\top]=\e^{G_n}\begin{pmatrix}
1 & 0 & 0& \cdots & 0
\end{pmatrix}^\top.
\]

\clearpage
\bibliography{references}
\bibliographystyle{chicago}
\end{document}